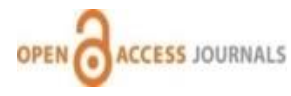

Research Article

# The Importance of the Clustering Model to Detect New Types of Intrusion in Data Traffic

**Author Details**
Noor Saud Abd[1],Noor Walid Khalid[2], Basim Hussein Ali[3]

**Authors Affiliations**
[1]Assistant Lecturer in the Department of Cyber Security, Computer Science and Mathematics College, Tikrit University – Iraq
[2]Assistant Lecturer in the Department of Computer Science, Computer Science and Mathematics College, Tikrit University – Iraq
[3]Sunni Endowment Diwan – Iraq

**Corresponding Author***
Noor Saud Abd
noor.s.abd@tu.edu.iq

**Abstract:** In the current digital age, the volume of data generated by various cyber activities has become enormous and is constantly increasing. The data may contain valuable insights that can be harnessed to improve cyber security measures. However, much of this data is unclassified and qualitative, which poses significant challenges to traditional analysis methods. In order to over- come these challenges, clustering, a crucial method in machine learning (ML) and data analysis, has become increasingly effective. Clustering facilitates the identification of hidden patterns and structures in data through grouping sim- ilar data points, which makes it simpler to identify and address threats. Clus- tering can be defined as a data mining (DM) approach, which uses similarity calculations for dividing a data set into several categories. Each data cluster that the clustering algorithm has identified has a high degree of similarity, and there is a fair amount of similarity between other clusters of data. Hierar- chical, density-based, along with partitioning clustering algorithms are typi- cal. The presented work use K-means algorithm, which is a popular clustering technique. Utilizing K-means algorithm, we worked with two different types of data: first, we gathered data with the use of XG-boost algorithm fol- lowing completing the aggregation with K-means algorithm. Data was gath- ered utilizing Kali Linux environment, cicflowmeter traffic, and Putty Soft- ware tools with the use of diverse and simple attacks. The concept could assist in identifying new attack types, which are distinct from the known attacks, and labeling them based on the characteristics they will exhibit, as the dy- namic nature regarding cyber threats means that new attack types of- ten emerge, for which labeled data might not yet exist. The model counted the attacks and assigned numbers to each one of them. Secondly, We tried the same work on the ready data inside the Kaggle repository called (Intrusion Detection in Internet of Things Network), and the clustering model worked well and detected the number of attacks correctly as shown in the results sec- tion.

## INTRODUCTION

With the world becoming more interconnected, there is a greater risk of cyber attacks and intrusions than before. For example, on October 21, 2016, numerous datacenters were struck by severe distributed denial of service (DDoS) attacks, which resulted in the closure of Spotify, Twitter, and other popular websites [1]. Over the years, there have been more threats to the integrity and confidentiality of data and systems along with denial-of-service (DoS) attacks [2].Modern cyber security strategies must include intrusion detection, which is responsible for spotting malicious activity, unauthorized access, and policy violations in a network or system [3]. Large amounts of labeled data are needed for training accurate models in supervised learning techniques, which are frequently used in conventional intrusion detection systems (IDS). But getting labeled data in the cyber security challenging is especially hard [4].

Labeling network traffic or system logs involves significant manual effort, expert knowledge, and time. Moreover, the dynamic nature of cyber threats means that new types of attacks frequently emerge, for which labeled data may not yet exist [5]. In response to these challenges, the use of unlabeled datasets in intrusion detection has gained increasing attention. Unlabeled datasets, which consist of raw, unannotated data, represent the vast majority of data available in real-world cyber security scenarios [6].These datasets capture the natural flow of network traffic, system events, and user behaviors, including both normal and potentially malicious activities [7]. The absence of labels in these datasets poses a unique challenge, but it also presents an opportunity: by applying advanced unsupervised learning techniques, such as anomaly detection, clustering, and more recently, self-supervised learning; it is possible to uncover hidden patterns as well as anomalies which might specify scurity breaches or emerging threats [8].

The exploration of unlabeled datasets in intrusion detection opens up new avenues for developing adaptive, scalable, and robust security solutions [9]. These methods can learn to detect unknown or novel attacks without relying on prior knowledge, making them particularly valuable in the ever-evolving landscape of cyber security [10]. As such, the study and application of tech- niques that can effectively utilize unlabeled datasets are crucial for advancing the field of intrusion detection as well as improving the overall security pos- ture of organizations [11].

The workflow regarding the unsupervised and supervised models is shown in Fig 1 [12]. The supervised model's workflow is comprised of multiple steps, as seen in Fig 1(A): dataset evaluation, data acquisition, model training, and optimization. While unsupervised models are utilized with an unlabeled da- taset, as demonstrated in Fig 1(B), supervised models need labeled data, ne- cessitating the application of more data evaluation methods. Following using optimization approaches, unsupervised models—in this case, K-means and principal component analysis (PCA) are assessed on an unknown data pattern. The Results section provides an overview of the methods and materials used.

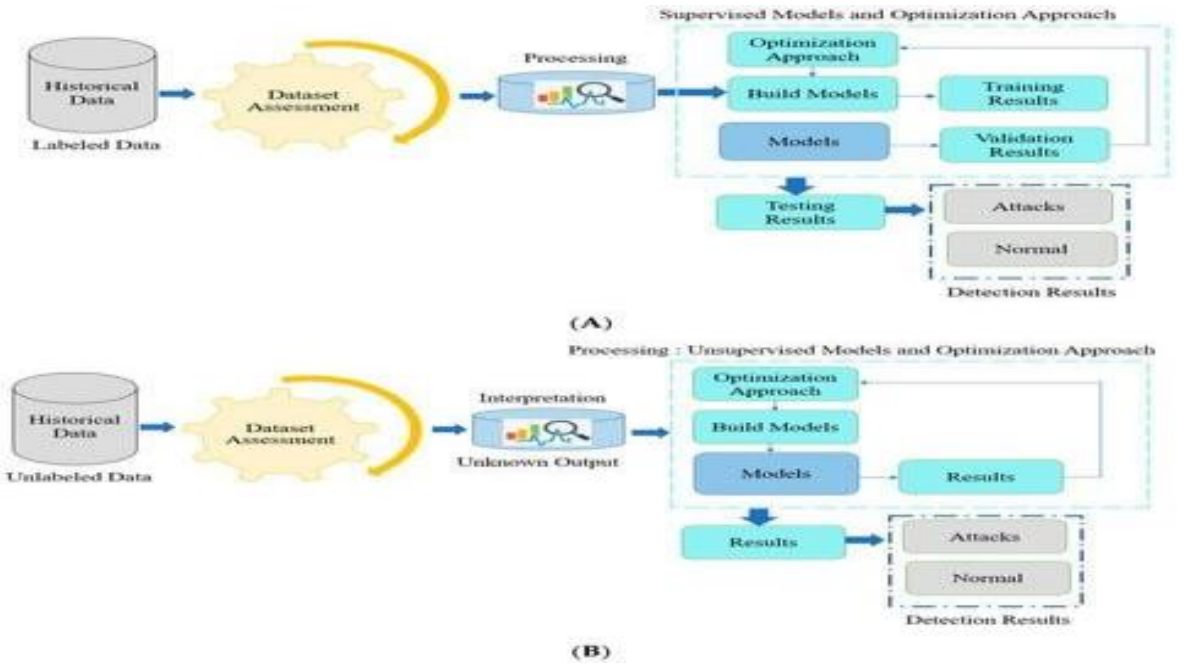

**(A)** Supervised Working Flow (B) Unsupervised Working Flow. Figure 1. Supervised and Unsupervised Learning Working Flow [12].

**Contributions of this paper:**

The clustering method was used here to find new types of attacks in case new types of them appeared, as we assumed that we do not know the names of the attacks present in a certain dataset (in our work we tried 2 datasets) and applied the clustering model to them and the model correctly identified the existing categories.

We created a dataset for simple and common types of attacks using the Kali Linux system and Putty and Cicflowmeter tools to record traffic (CICFlow- Meter is a network traffic flow generator and analyzer distributed by the Ca- nadian Institute for Cybersecurity (CIC). It is an open-source application writ- ten in Java and can be downloaded from Github. We used it to read the pcap files generated by tcpdump and provide csv files.), there were 7 types of at- tacks and the seven attacks were actually classified by the clustering system using the K-means algorithm.

While the other datasets were ready as indicated in the Dataset and Resources section at the end of the paper, to ensure the validity and robustness of the model.

We show related works in Section 2. Section 3 describes the 3 Clustering Technique and K-means clustering algorithms, while Section 4 describes the Methodology then the results. Finally, we wrap up our research in Section 5.

## 1.1 Related Work

With regard to such context, IDSs are being developed more and more daily so that the network systems could be effective against the developed malware. Thus, there are numerous literature studies as well as new studies are performed daily for increasing the performance regarding IDS systems. Attackers are constantly creating new attack scenarios and updating them- selves and the software they use.

- **Dhaliwal, S. S., Nahid, A. A, 2018 [13],** this study presents a model intended for measuring multiple network data attributes, including ac- curacy, precision, confusion matrix, and so on. To get the desired out- comes, XGBoost is applied to NSL-KDD (Network Socket Layer Knowledge Discovery in Databases) dataset. The primary goal is to identify the data's integrity and increase prediction accuracy. The like- lihood of data being hacked or altered decreases with network security. The researchers stress that through adjusting different model parame- ters, further research could be done in the future to maximize the amount of data entering and leaving the network. The data is the most significant participant in the network, and knowing it well and accu- rately is half the job done. An effective IDS, which preserves the net-work's integrity and provides a secure environment for exchanging con- fidential data, emerges as a result of studying the data in the network and evaluating its volume and pattern.

- **Rosay, A., Carlier, F., 2021 [14],** Many datasets were created during the last few decades to solve this security issue. Analysis of earlier datasets, like KDD-Cup99 and NSL-KDD, has brought attention to some of the difficulties and opened the door for the correction of those faults in more recent datasets. One of the most recent network intrusion detection da- tasets, CIC-IDS2017, is thoroughly analyzed in this research. They pre- sent a number of problems they found in the network packet flows which were recovered. A novel feature extraction technique named Ly- coSTand is suggested as a solution to these problems. In order to com- prehend network traffic, a tool called LycoSTand models network flows and extracts attributes that define them. The Greek word for wolf is lycos, which means flow in reverse order. Furthermore, a feature selection ap- proach is suggested that considers feature importance and account rela- tions. When comparing the new and original datasets' performances, all of the ML methods under evaluation exhibit significant improvements.They looked at other datasets impacted by the same problems that Ly- coSTand may be utilized to solve in order to create better datasets for network intrusion detection, depending on the improvements in CIC- IDS2017.
- **Farhan, R. I., Maolood, (2020) [15],** this work suggests solving the fea- ture selection (FS) problem with binary Particle Swarm Optimization (BPSO). Next, using the CSE-CIC-IDS2018 dataset as well as Deep Neural Networks (DNN) classifiers, features chosen from BPSO are as- sessed. When compared to other benchmark classifiers, the suggested model's performance has demonstrated similar results in terms of pro- cessing time, false alarm rate, and detection rate. The experimental find- ings indicate a 95% accuracy rate.
- **Silivery, A. K., Rao, K. R. M., (2023) [19],** this research develops an efficient DL-based IDS for the classification of DDoS and DoS attacks. In order to solve the problem of class imbalance in the dataset, the model consists of multiple steps, beginning with an approach based on Deep Convolutional Generative Adversarial Networks (DCGAN). Next, the essential features for every class in the dataset are extracted using a DL algorithm built on top of ResNet-50. Subsequently, an optimized AlexNet-based classifier is constructed to identify the attacks individu- ally. Atom search optimization technique is utilized for optimizing the classifier's key parameters. With the use of important classification crite- ria, the suggested method was assessed on the benchmark datasets CCIDS2019 as well as UNSW-NB15. It obtained an accuracy of 99.37% for UNSW-NB15 dataset and 99.33% for the CICIDS2019 dataset. The investigational findings show that the recommended method outperforms competing approaches in detecting DDoS and DoS.
- **Sakthi, K., & Nirmal Kumar, P. (2023) [16],** this study research pre- sents an optimal weight-based deep neural network (OWDNN) based DL approach for network intrusion detection (NID). The first three publicly accessible datasets from which the network traffic data was gathered were CSE-CIC-IDS2018, NSL-KDD, and UNSW-NB15. The gathered data was after that preprocessed using one-hot encoding, normalization, and imputation for missing values. The butterfly-optimized k-means clustering (BOKMC) algorithm is then used to do the data under-sam- pling process for balancing the unbalanced dataset. To lighten the classi- fier's computational load, the inception version 3 with multi-head atten- tion (IV3MHA) technique is used to choose the relevant features from the balanced dataset. Next, PCA is used for reducing the dimensionality of the chosen feature. Lastly, OWDNN is used to classify the network traffic, dividing it into normal and anomalous categories. Tests con- ducted on the CSE-CIC-IDS2018, NSL-KDD, and UNSW-NB15 da- tasets demonstrate that OWDNN outperforms the other ID techniques.

### 1.3 Clustering Technique

A key method in data analysis is clustering, which groups a set of objects so that those in the same group referred to as a cluster are more similar to one another compared to those in other groups. In the field of cyber security, this method has grown in importance, especially when working with qualitative and unclassified data [17]. Putting data into clusters depending on similarity is the process of clustering, an unsupervised learning technique. Data that has been labeled is not necessary for clustering, in contrast to classification, when data points are given predefined labels. Rather, it aims to divide the dataset into clusters whose members are more alike than they are from other clusters. Unlike classification, clustering allows for model modifications and the creation of sub-clusters. Because of this, clustering is especially helpful in situations where the underlying data structure is unclear and in exploratory data analysis [18]. Numerous clustering techniques exist, and each is appropriate for a particular set of data and use cases. Several of the typical methods consist of:

1. K-means Clustering: Each data point belongs to the cluster with the nearest mean in this algorithm, which divides the data into K clus- ters. Because of its efficiency and simplicity, it is widely utilized.
2. Hierarchical Clustering: Using a top-down (divisive) or bottom-up (agglomerative) method, this approach creates a hierarchy of clus- ters. It is helpful for organizing clusters into a structure resembling a tree.
3. DBSCAN (Density-Based Spatial Clustering of Applications with Noise): This algorithm classifies as outliers the points that are iso- lated in low-density regions and clusters together points that are densely packed. It works well for locating clusters of any shape.
4. Spectral Clustering: This method reduces dimensionality and per- forms clustering in fewer dimensions by utilizing the eigenvalues regarding a similarity matrix. It is especially helpful for clusters that are not convex.
5. Gaussian Mixture Models (GMM): assumes that a mix of many Gaussian distributions with varying parameters produced the data.

Lastly, grouping related objects into the same class without label information is referred to as clustering. These days, cluster analysis is a popular unsuper- vised technique that is also frequently utilized in DM. Due to its straightfor- ward concept, concise algorithm, and strong clustering impact, the K-means clustering algorithm has garnered a lot of interest from scholars and has been used in a variety of domains [19].

### 1.3.1 Applications of Clustering in Cyber security

In the context of cyber security, clustering helps in various ways, especially when dealing with qualitative and unclassified data [20]:

- Intrusion Detection: Clustering can identify abnormal patterns or out- liers in network traffic that might indicate security breaches or intru- sions. By analyzing logs and traffic data, clustering helps in distin- guishing between normal and anomalous behaviors. Malware Detection: By clustering different types of malware based on their behavior or code similarities, cyber security professionals can identify new malware variants and understand their potential impact.
- Threat Intelligence: Clustering threat data (e.g., IP addresses, attack signatures) enables the identification of related threats and attack pat- terns. This helps in predicting future attacks and preparing defenses accordingly.
- Phishing Detection: Clustering techniques can help in identifying phishing attempts by analyzing email content, URLs, and sender infor- mation. Clustering similar phishing emails together makes it easier to detect new phishing campaigns.

Finding data items with comparable characteristics and understanding the dif- ferences and similarities between variables are made easier by the clustering process. Classification and clustering are similar; however clustering does not enable the instantaneous organization of variables. If you select this method, you will only receive help with organizing and analyzing an already-existing database. Unlike classification, clustering allows for model modifications and the creation of sub-clusters [21]. In our study, we employed the J.B. MacQueen-proposed K-Means algorithm, which divides data into groups. This unsupervised algorithm is typically employed in pattern recognition and DM applications. The cornerstones of this algorithm are error criterion, square error, and cluster performance index minimization. This algorithm seeks K divisions that satisfy a specific criterion in order to find the optimal result [22]. First, select a few dots to represent the initial cluster focal points (typically, first K sample dots of income); next, group the mining sample dots to their focal points based on the minimum distance criterion. This will yield the initial classification; if it is deemed unreasonable, we will revise it (recal- culate each cluster focal point again), and this process will be repeated until we obtain a reasonable classification. The division-based means algorithm is a type of cluster algorithm that offers the benefits of speed and effi- ciency [23]. Nevertheless, the algorithm's results are constantly dependent on the initial dots and the differences in beginning sample selection. Further- more, the gradient approach is often used by this target function-based algo- rithm to obtain the extremum. In the case when the initial cluster focal point is improper, the gradient approach's search direction is always along the di- rection of energy decline, which makes it easy for the entire algorithm to sink into the local minimum point [24].

### 1.3.2 K-means clustering algorithms

The unlabeled dataset is grouped into various clusters using the K- Means clustering algorithm, which is an Unsupervised Learning technique [25]. In this case, K indicates how many pre-defined clusters must be formed during the process; for example, if K=2, there will be two clusters; if K=3, there will be three clusters, and so on. Still the most often used and simple clustering algorithm is the K-means algorithm. The algorithm's low compu- ting complexity and ease of implementation account for its broad applicability across numerous clustering application domains. Nevertheless, a number of challenges that the K-means algorithm has have a detrimental impact on how well it clusters data [26]. Clustering is an effective unsupervised learning method that divides unknown things into many groups. Every group's mem- bers share comparable characteristics and attributes. One comparatively easy method for implementing clustering analysis is K-means cluster algorithm [27]. The authors of the algorithm, have provided comprehensive details. Its principle is as fol- lows:

**i-** The input data are first split into many groups after a predetermined num- ber K of initial centroids are randomly determined. The following formula is used to get the Euclidean distance (d) between the centroids and the data [28]:

$$d(X_t, X_\xi) = \sqrt{\sum_{u=1}^{n} (X_{ut} - X_{u\xi})^2} \quad (1)$$

In which, u represent the data property, l represent the number of properties, and Xt and Xξ are the input data and supplied centroids, respectively. In the presented work, n is fixed to 1 since the occurrence probability of the land- slide serves as the data property for modeling the susceptibility to landslides.

ii. The new centroids are updated with the use of the next equation once the first computation of d is complete, and the result is noted as D (0)={D1 (0), D2(0),..., DK (0)}

$$X_\xi^{(m)} = \frac{1}{h_\xi^{(m-1)}} \sum_{X_t = D_{t\xi}^{(m-1)}} X_t \quad (2)$$

where Xξm represent the new centroid, m represents the number of itera- tions, hξ(m−1) represent the amount of data in the new group depending on new centroids.

ii. iterating the above step, and ending the calculation process when ξ(m) = hξ(m−1) and D(m) = D(m−1). The obtained centroids are the clustering centers of the input data [29].

## 1.4 METHODOLOGY

Generating unclassified cyber-attack data in Kali Linux can be useful for learning, testing, and developing cyber security tools and techniques. There are some steps to simulate and capture attack data using various tools in Kali Linux, these tools and techniques are: 1- Metasploit Frame- work 2- Wireshark 3- tcpdump 4- hping3. In our experiment we used tcpdump is a command line packet analysis tool. As we show in steps below:

1- Install tcpdump: sudo apt update

sudo apt install tcpdump 2- Capture Traffic:

sudo tcpdump -i <interface> -w attack_data.pcap

Replace <interface> with your network interface (e.g., eth0, wlan0).

3- Stop Capture:

Press Ctrl+C to stop capturing traffic. Dataset

We planned and deployed two networks, the Attack-Network and the Victim- Network, in order to build a thorough tested. As we previously indicated, we attempted to create a dataset that is unique to the Internet. We worked with Kali Linux and the tcpdump program with Putty to acquire data that was 19,677 rows and 84 columns. However, the data needs to be properly prepared before being used in the model.

Data preprocessing has three steps:

- Data Cleaning: specify missing values as well as fix errors.
- Data Digitization: converting categorical features into numerical val- ues.
- Data Transforms: like normalization for changing scale, type, and probability distribution regarding variables in the dataset.

After working on Kali Linux and the tcpdump tool using PuTTY, we obtained data that was (19,677 rows, 83 columns), and the figure (1) shows a sample of it:

There is a small percentage of missing values in the collected data set, and a drop was made for the column containing these missing values, as well as for some values that are not important in later processing, which are (Flow ID", "Src IP", "Dst IP", "Timestamp"), so it became Dataset (19677 rows, 78 columns).

We did not notice missing values, except for except for a few values in one column, and we dropped it without affecting the quality of the data and the work of the algorithms.

We also made a drop for a set of data that is not important in the classi- fication process, namely ("Flow ID", "Src IP", "Dst IP", "Timestamp"). Create a new column in the DataFrame to store the cluster labels scaled_df['Cluster_Labels'] = cluster_labels scaled_df["Cluster_Labels"].value_counts()

Cluster_Labels 1 7361

0 5801

2 3390

3 1349

4 1080

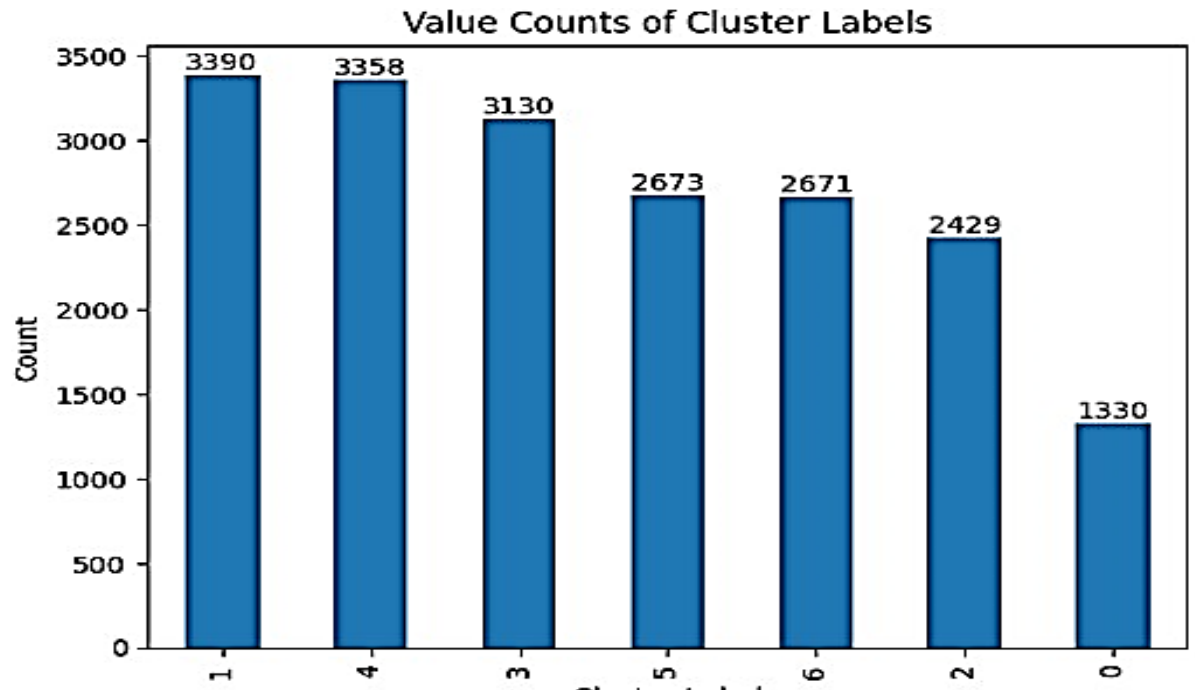

**Figure (2)** the Distribution of the Classes

**Dataset Description**

Two datasets were used in this paper: the IoT Intrusion Detection da- taset from the Kaggle repository. And a dataset created in an actual test implementation using devices that generate real traffic (including DoS, Brute Force Attack, TCP Floading, UDP Floading, etc.) that simulates user behavior. Subsets of their normal traces were included in the train- ing dataset as well as a subset of normal traffic in the test dataset.

A dataset obtained from tcpdump captures raw network traffic data, representing packets transmitted over a network during the capture pe- riod. This dataset is usually stored in a. pcap (Packet Capture) file for- mat. Below is a detailed description of the typical contents and struc- ture of a tcpdump dataset:

1- Packet Metadata:

- Timestamp: The exact date and time in the case when the packet has been captured. This is important for analyzing the sequence and timing of events in network traffic.
- Frame Number: A unique identifier allocated to each packet in cap- ture, useful for analysis and reference.

2- Ethernet Header:

- Source MAC Address: The hardware address of the device that sent the packet.
- Destination MAC Address: The hardware address of the device that is intended to receive the packet.
- EtherType: Indicates the protocol of the encapsulated data, such as IPv4, ARP, or IPv6.

3- Network Layer (IP Header):

- Source IP Address: The IP address of the device that sent the packet.
- Destination IP Address: The IP address of the device that is intended to re ceive the packet.
- Protocol: Indicates the protocol used in the data portion of the packet (e.g., TCP, UDP, ICMP).

4- Transport Layer:

- Source Port: The port number on the source device from which the packet was sent.
- Destination Port: The port number on the destination device to which the packet is addressed.
- Sequence Number: Used in TCP connections to ensure packets are received in order.
- Acknowledgment Number: Used in TCP connections to confirm re- ceipt of packets.
- Flags: Control bits that manage the state of the TCP connection (e.g., SYN, ACK, FIN).

5- Application Layer:

Payload Data: The actual data being transmitted by the application. This could be HTTP requests/responses, DNS queries/responses, or any other protocol data.

**ii- Data Preparation:**

There are steps for data preparation that we list with a simple explana- tion for each one.

**A. Data Collection:**

After working on Kali Linux and the tcpdump tool using PuTTY, we obtained data that was (19,677 rows, 83 columns)

**B. Data Cleaning:**

Remove or handle missing values, outliers, and irrelevant features. Clean and preprocess the data to ensure quality input for the model. In our dataset, there was not much missing data except for one column that was dropped it without affecting the quality of the data and the work of the algorithms.
There is a small percentage of missing values in the collected data set, and a drop was made for the column containing these missing values, as well as for some values that are not important in later processing, which are (Flow ID", "Src IP", "Dst IP", "Timestamp"), so it became Dataset (19677 rows, 78 columns).

We also made a drop for a set of data that is not important in the clas- sification process, namely ("Flow ID", "Src IP", "Dst IP", "Timestamp").

**C. Data Splitting:**

Divide the dataset into sets for testing and training. To assess the per- formance of the model, a typical split may be 20% for testing and 80% for training.

### 1.1.2 Building the model

**i - Building the XGBoost Model**

An effective and powerful of gradient boosting algorithms built for perfor- mance and speed is known as XGBoost, or Extreme Gradient Boosting Algo- rithm. Because of its accuracy, scalability, and flexibility, it has gained pop- ularity as a choice for developing predictive models. XGBoost has a wide range of applications in cybersecurity, including malware classification, in- trusion detection, and anomaly detection [30]. First, specify the XGBoost model's hyperparameters. With the use of training data, train the XGBoost model. Make predictions on test data using the trained model. The next fac- tors explain why XGBoost is specifically chosen as the best classification model to address problems that arise in real word classification tasks:

- One platform means that time consumption will be eliminated, particu- larly in the case when pre-processing network data.

XGBoost benefits from parallel processing, which makes use of all the cores on the computer it is operating on. It is extremely scalable, uses few resources, and produces billions of examples through algorithmic optimization pro- cesses and distributed or parallel computing. As a result, it works incredibly well when handling problems like data classification and advanced pre-pro- cessing.

XGBoost is more accessible and simpler to integrate across a variety of plat- forms due to its portability. The dispersed versions are now being linked with cloud computing platforms including Alibaba's Tianchi, GCE, AWS, Azure, and more. As a result, XGBoost offers tremendous flexibility that is not platform-specific, making it possible for IDS that use it to be platform-independ- ent, which is a significant benefit [31]. Additionally, XGBoost is interfaced with Flink and Spark, two cloud data flow systems.

A variety of programming languages, including Python, Java, R, and C++, could handle XGBoost.

- XGBoost could be utilized in a variety of computing scenarios, such as out- of-core computing, parallelization (tree construction across many CPU Cores), distributed computing for handling large models, and cache optimi- zation for efficient hardware use.

The capacity of XGBoost to optimize each new tree it attaches, turning a weak learner into a strong learner (boosting); this enables the classification model to produce fewer False Alarms and facilitates accurate and simple data clas- sification.

Regularization is a key component of XGBoost algorithm since it prevents problems with data overfitting in both tree-based and linear models. In the case when a system is experiencing a DDoS attack, or flooding of data entries, XGBoost could effectively handle data-overfitting problems. As a result, the classifier must be quick—which XGBoost is—and flexible enough to handle data entries.

- Cross-validation is activated as an internal function. As a result, extra pack- ages are not required to obtain the results of cross validation.
- XGBoost has the tools necessary to identify and handle missing values.
- XGBoost is a flexible classifier since it allows the user to customize the goal function by adjusting the model's parameters. It handles classification, regres- sion, and ranking problems in addition to supporting user-defined evaluation metrics.
- XGBoost is simple to use and accessible because to its availability across various platforms.
- The data matrix can be saved and relaunched when needed, thanks to XGBoost's save and reload functions. This removes the requirement for more RAM.
- Extended Tree Pruning: Under standard models, tree pruning ends as soon as a negative loss is observed. However, with XGBoost, pruning continues up to a user-specified maximum tree depth after which backward pruning is car- ried out on the same tree until the improvement in the loss function falls below a predetermined threshold value [32].

The combination of such significant features allows the XGBoost to surpass numerous current models. Furthermore, XGBoost was utilized in a large number of winning entries in ML contests like KDDCup (all 10 winning en- tries in 2015) and Kaggle (17 out of 29 winning entries in 2015). As a result, it could be concluded that XGBoost is highly capable of handling the vast majority of issues that arise in real-world networks [38, 39]. A well-liked ML algorithm called XGBoost is mostly applied to supervised tasks like clas- sification and regression. Yet, in the case when paired with unsupervised methods, it could be modified for tasks such as clustering. It is rare to use XGBoost directly for clustering in cyber intrusion detection, however it could be incorporated into a hybrid model. Allow me to explain how this might operate [33]:

**A-** XG-Boost for Classification or Anomaly Detection:

XGBoost is typically used for classification in intrusion detection, where it distinguishes between normal and anomalous activities based on labeled data. Here's how it can contribute to intrusion detection:

- Input Features: Features extracted from network traffic data (like IP ad- dresses, packet sizes, timing info, etc.).
- Target Labels: The labels in supervised settings (e.g., "benign" or "intru- sion").
- Modeling: XGBoost creates an ensemble of decision trees that try to pre- dict whether a given instance is normal or an intrusion. It does this by boosting the accuracy of weak classifiers.

**B-** Here are a few ways clustering is integrated:

- Feature Engineering for XGBoost: Clustering algorithms (like K-Means, DBSCAN, etc.) can be used before running XGBoost to group similar pat- terns. These cluster labels (or distances from cluster centers) can be added as features to train the XGBoost model. This adds an extra layer of struc- ture to the data, helping XGBoost differentiate between normal and ab- normal traffic.
- Anomaly Detection Using Clustering and XGBoost:
  - Step 1: Perform clustering on the cyber intrusion data to find naturally occurring groups in the data (e.g., normal traffic vs. anomalous traffic).
  - Step 2: Use the clustering results as labels for semi-supervised training.
  - Step 3: Apply XGBoost to refine the detection of

anomalies by learning from these clusters, identifying subtle deviations in the data.

**C-** Hybrid Approach (Clustering + XGBoost):

You can combine clustering and XGBoost in several ways:

- Pre-processing with Clustering: Use clustering (like K-means) to generate pseudo-labels and train XGBoost using these clusters.
- Post-clustering Prediction: After clustering to detect anomalies, XGBoost can be used to classify the anomalies into different categories or refine the results.

**D-** Workflow for Intrusion Detection with Clustering + XGBoost:

1. Data Collection: Collect network data, including logs and packet captures.
2. Clustering: Apply a clustering algorithm (e.g., DBSCAN) to detect poten- tial anomalous clusters in the traffic data.
3. Labeling: Use the cluster labels as a form of pseudo-labeling or initial anomaly detection.
4. XGBoost Training: Train an XGBoost classifier using these clusters (or distances from the clusters) as features for more refined detection.
5. Intrusion Prediction: Use the trained XGBoost model to classify new in- coming data as normal or potential intrusions.

**E-** Challenges in Cyber Intrusion Detection:

- High-dimensional Data: Cybersecurity datasets can be large and complex, but XGBoost is well-suited for handling large datasets with many features.
- Imbalanced Data: Intrusion detection datasets are often imbalanced, meaning attacks are rare compared to normal traffic. XGBoost handles this well through its ability to focus on harder-to-classify examples.

**ii- Performance Evaluation:**

Evaluate model performance using metrics such as accuracy, confusion ma- trix, and classification report.

**Table 1. Classification Report**

| The name of attack | Precision | Recall | F1-score | Support | Accuracy |
|---|---|---|---|---|---|
| 0 | 0.99 | 0.98 | 0.98 | 252 | 99.9 |
| 1 | 1.00 | 1.00 | 1.00 | 678 | 1.00 |
| 2 | 1.00 | 1.00 | 1.00 | 498 | 99.9 |
| 3 | 0.99 | 1.00 | 0.99 | 630 | 99 |
| 4 | 1.00 | 1.00 | 100 | 646 | 1.00 |
| 5 | 0.99 | 1.00 | 1.00 | 539 | 99.8 |
| 6 | 0.99 | 1.00 | 0.99 | 554 | 99.7 |

Cohen's Kappa Score: 0.9947234039909504

Precision: 1.0

Recall: 1.0

NPV: 1.0

PPV: 1.0

Sensitivity: 1.0

Specificity: 1.0

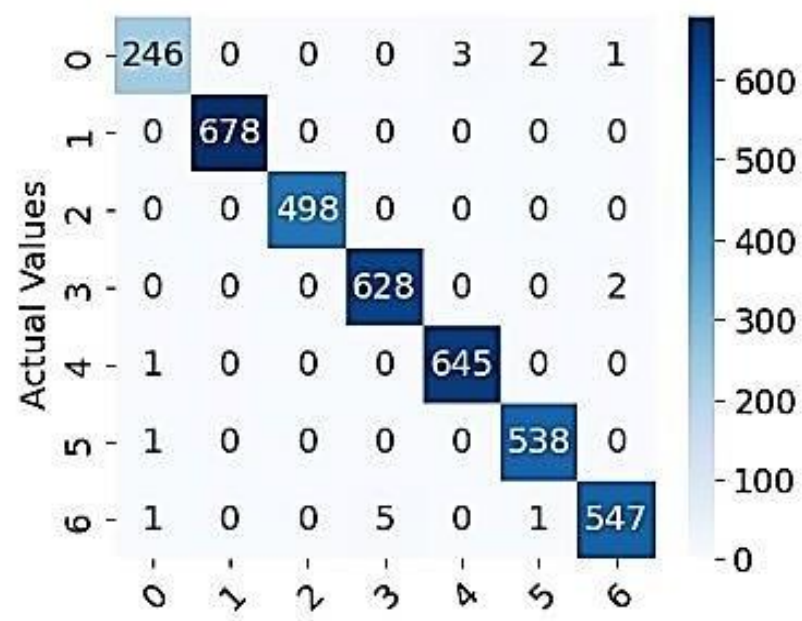

**A- Confusion Matrix**

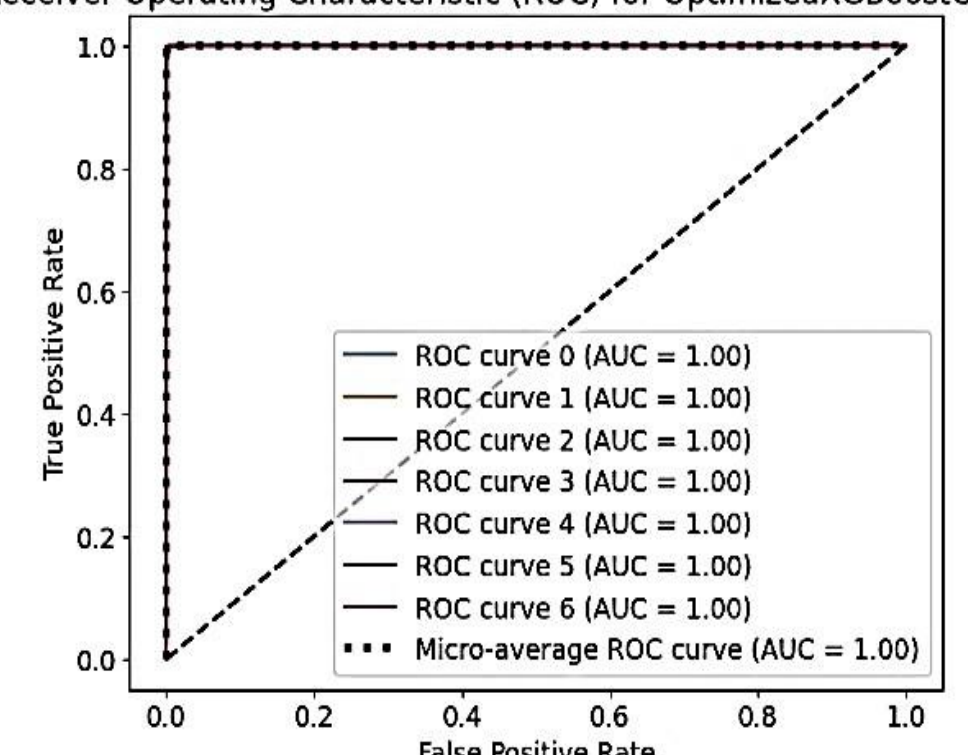

B- Roc Curve Figure (3) Confusion Matrix and Roc

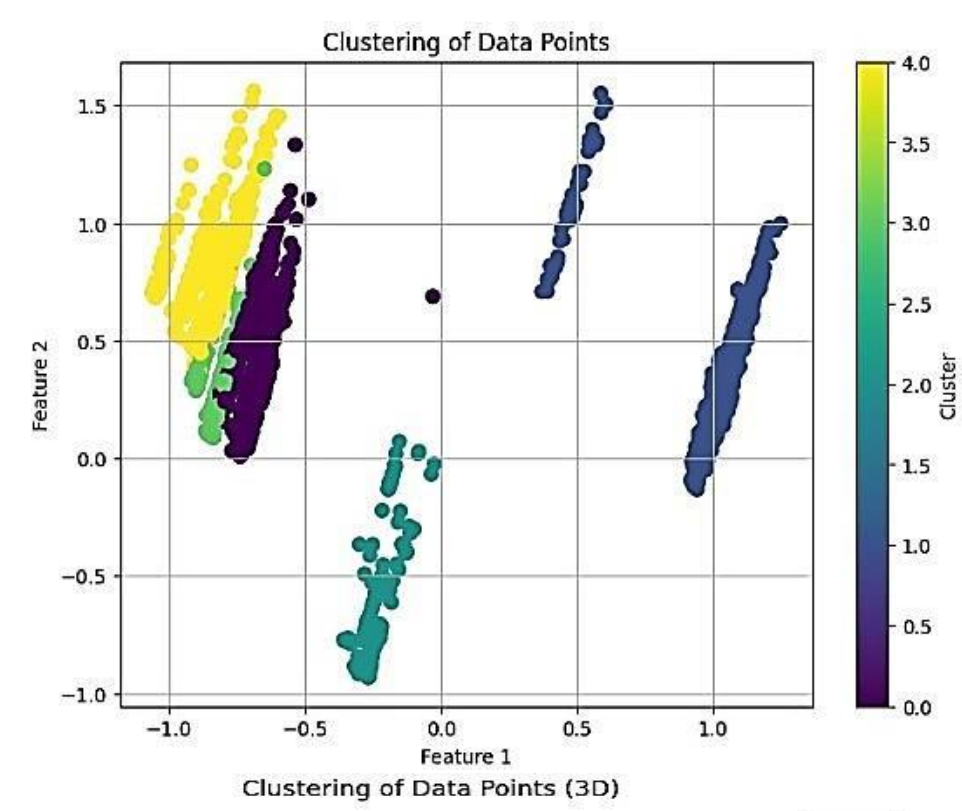

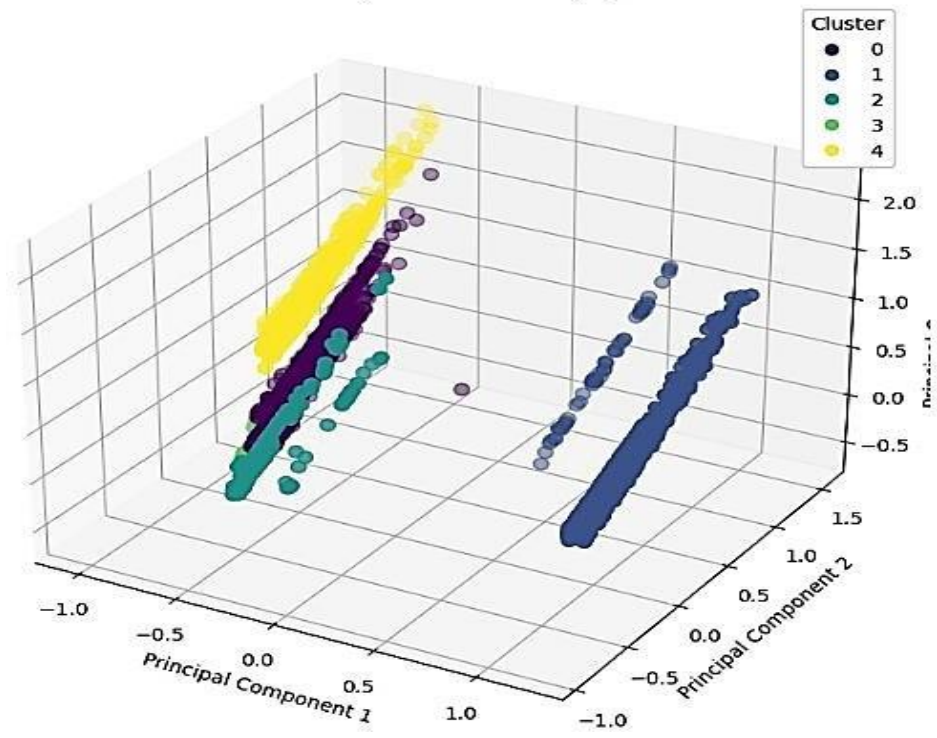

Figure (4) Some Visualization about the Distribution of the Classes( Each color represents a specific type)

## 1.2 CONCLUSION

Availability of datasets is one of the biggest challenges in the field of intru- sion detection systems. Due to privacy and security reasons, most organiza- tions will never share their network traffic data. However, a high-quality da- taset is crucial to develop an anomaly-based intrusion detection system and evaluate its performance. Therefore, in this research, we have tried to perform simple common attacks at close times and test them on the proposed model if it can recognize the presence of a group of unclassified attacks and group them into groups, then pass them to the classification process and complete the rest of the requirements such as knowing the accuracy, confusion matrix, etc.

Clustering is a versatile and powerful tool in the arsenal of cyber security professionals. Its ability to handle unclassified and qualitative data, uncover hidden pat-terns, and detect anomalies makes it indispensable in the ongoing battle against cyber threats. As technology continues to evolve, clustering techniques will undoubtedly become even more integral to maintaining robust cyber security defenses, providing deeper insights and more proactive protec- tion against an ever-expanding array of cyber threats.

### Acknowledgements

The authors would like to acknowledge the Computer Science Department, Tikreet University and the National Institute of Applied Sciences and Tech- nology - Tunisia.

### Funding

This research received no specific grant from any funding agency in the pub- lic, commercial, or not-for-profit sectors.

**Conflict of Interest:** The authors declare that they have no conflict of interest

**Funding:** No funding sources

**Ethical approval:** The study was approved by the Tikrit University – Iraq